\documentstyle[epsfig]{aipproc}

\begin{document}
\title{SNEWS: The SuperNova Early Warning System}

\author{Kate Scholberg$^*$}
\address{$^*$Boston University Department of Physics, 590 Commonwealth Ave., Boston, MA 02215}

\maketitle

\begin{abstract}
World-wide, several detectors currently running or nearing completion
are sensitive to a prompt core collapse supernova neutrino signal in
the Galaxy.  The SNEWS system will be able to provide a robust early
warning of a supernova's occurrence to the astronomical community
using a coincidence of neutrino signals around the world.  This talk
describes the nature of the neutrino signal, detection techniques and
the motivation for a coincidence alert.  It describes the
implementation of SNEWS, its current status, and its future, which can
include gravitational wave detectors.
\end{abstract}

\section*{The Expected Neutrino Signal}

When the core of a massive star at the end of its life collapses,
nearly all of the total gravitational binding energy of a neutron star
is emitted in the form of neutrinos, some
$E_b~\sim~3\times10^{53}$~ergs.  Less than 1\% of this energy is
expected to be released in the form of kinetic energy and optically visible
radiation. The remainder is radiated in
neutrinos, of which approximately 1\% will be electron neutrinos from
an initial ``neutronization'' burst and the remaining 99\% will be
neutrinos from the later cooling reactions, equally distributed among
flavors.  Average neutrino energies are expected to be about 12 MeV
for electron neutrinos, 15 MeV for electron antineutrinos, and 18 MeV
for all other flavors.  The neutrinos are emitted over a total
timescale of tens of seconds, with about half emitted during the first
1-2~seconds, and with the spectrum eventually softening as the
proto-neutron star cools.  Reference~\cite{Burrows} summarizes the
expected neutrino signal.  The basic features of neutrino emission
models were well confirmed in 1987A with the observation of neutrinos
from SN1987A.  We await the next Galactic supernova to learn more.

\section*{Neutrino Detectors}\label{detectors}

There are several classes of detectors capable of detecting a burst of
neutrinos from a gravitational collapse in our Galaxy.
Table~\ref{tab:detector_types} gives a brief overview; more details
can be found via reference~\cite{snews}.
Table~\ref{tab:specific_detectors} lists some specific supernova
neutrino detectors and their capabilities.

\begin{table}[t]
\caption{Supernova neutrino detector types.}
\label{tab:detector_types}
\begin{tabular}{||c|c|c|c|c|c||}\hline\hline
Detector type & Material & Energy & Time & Point & Flavor \\ \hline\hline
scintillator & C,H & y & y & n & $\bar{\nu}_e$\\ \hline
water Cherenkov &  H$_2$0 & y & y & y & $\bar{\nu}_e$ \\ \hline
heavy water & D$_2$0 & NC: n & y & n & all \\ \cline{3-6}
         &      & CC: y & y & y & $\nu_e$,$\bar{\nu}_e$\\ \hline
long string water Cherenkov& H$_2$O & n & y & n & $\bar{\nu}_e$ \\ \hline
liquid argon & Ar & y & y & y & $\nu_e$ \\ \hline
high Z/neutron  & NaCl, Pb, Fe & n & y & n & all \\ \hline
radio-chemical & $^{37}$Cl, $^{127}$I, $^{71}$Ga  & n & n & n & $\nu_e$ \\
 \hline \hline 
\end{tabular}
\end{table}


\begin{table}[b]
\caption{Specific supernova neutrino detectors.} 
\label{tab:specific_detectors}
\begin{tabular}{||c|c|c|c|c|c||}\hline\hline
Detector & Type & Mass & Location  & \# of events & Status\\ 
         &      & (kton) &         & @8.5 kpc &  \\ \hline\hline
Super-K & H$_2$O Ch. & 32    & Japan  & 5000 & online \\ \hline
MACRO & scint.& 0.6 & Italy & 150 &  online \\ \hline
SNO & H$_2$O,& 1.4 & Canada &  300 & running \\
    & D$_2$O & 1   &        &  450 &  \\ \hline
LVD & scint. & 0.7 & Italy & 170 & online\\ \hline
AMANDA & long string  & M${\rm eff}\sim$0.1/pmt & Antarctica & & running \\ \hline 
Baksan & scint. & 0.33  & Russia  & 50 &  running\\ \hline
Borexino & scint. & 0.3  & Italy  & 100 &  2001\\ \hline
KamLAND & scint. & 1  &  Japan  & 300 &  2001\\ \hline
OMNIS (Pb/Fe) & high Z   & 5 &  USA & 2000 &  2000+ \\ \hline
LAND (Pb) & high Z   & &  Canada &  &  2000+ \\ \hline
Icanoe & liquid argon & 9 & Italy & & 2000+ \\ \hline\hline
\end{tabular}
\end{table}

\section*{Early Supernova Observation}

The neutrino burst produced by the core collapse emerges promptly from
the stellar envelope.  However, the the shock wave produced by the
collapse takes some time to travel outwards from the core to the
photosphere of the star.  The time of first shock breakout of a
supernova is highly dependent on the nature of the stellar envelope,
and can range from minutes for bare-core stars to hours for red
giants.  For SN1987A, first light was observed about 2.5 hours after
the neutrino burst; the first observable photons probably occurred
about one hour earlier than that.

The observation of very early light from a supernova just after shock
breakout is astrophysically very interesting~\cite{HST}, and
rare for extragalactic supernovae. The environment
immediately around the progenitor star is probed by the initial stages
of the supernova.  For example, any effects of a close binary
companion upon the blast would occur very soon.  In addition, shock
breakout may also be accompanied by a UV and soft x-ray flash.  The
tail of such a flash was observed by the EUVE satellite for SN1897A.
And of course, an observation of very early supernova light could also
yield entirely unexpected effects.  

It is possible that a core collapse event will not yield an
optically bright supernova, either because the explosion ``fizzles'',
or because the supernova is in an optically obscured region of
the sky.  In the latter case there may still be an observable
event in some wavelengths, or in gravitational radiation.

\section*{Coincidence of Neutrino Signals}

There are several benefits from a system which coordinates neutrino
signals from two or more different detectors.  All detectors are
subject to false alarms due to bursts of events due to detector
pathologies or other non-Poissonian phenomena (for example, flashing
phototubes or other sources of spurious light, electronic noise,
correlated radioactivity events due to muon spallation of nuclei,
etc.).  Therefore, if an individual experiment is to issue an alarm, a
human operator must first check the event burst to confirm its
supernova-like nature, which can take significant time even when a
fast-response human alert system is set up.  Requiring a coincidence
between independent detectors will add great confidence to the
detection of a supernova neutrino burst, to the extent that a
completely automated alert may be possible.  The automation could save
enough time that important early observations would not be lost.

\section*{The SNEWS System: Implementation of a Coincidence Monitor}

\begin{figure}[b!] 
\centerline{\epsfig{file=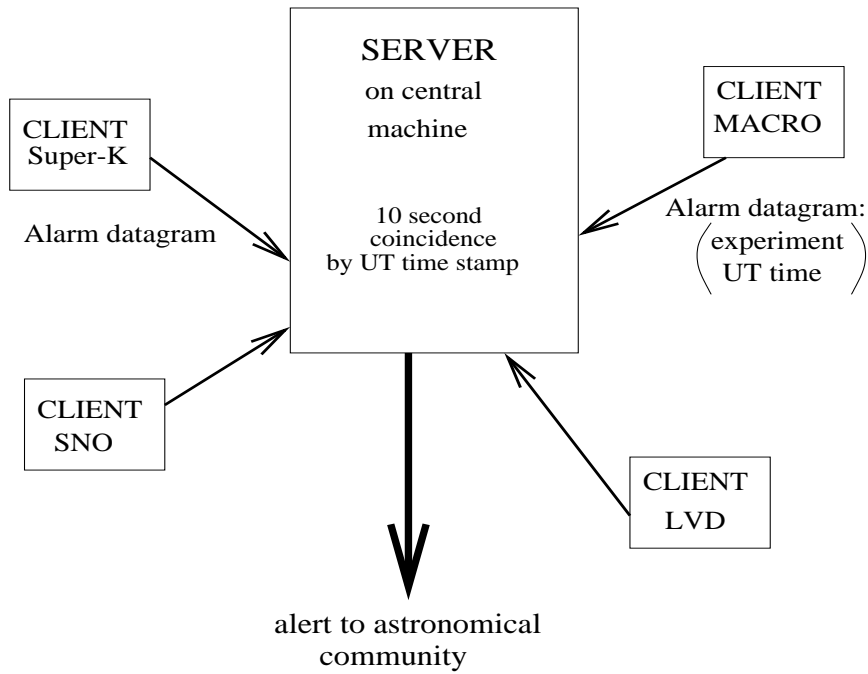,height=3.5in,width=4.5in}}
\vspace{10pt}
\caption{SNEWS setup.}
\label{fig:gcsetup}
\end{figure}

Software for a prototype international supernova watch coincidence
system has been designed by Alec Habig and Kate Scholberg.  It is
written in standard C and uses a standard UDP protocol client/server
setup to make direct network connections using sockets.  Dedicated
phone lines could be used to increase reliability if it proves necessary.
Figure~\ref{fig:gcsetup} shows the setup.

A central machine runs a ``server'' program, which sits and waits for
input from the outside.  The individual experiments participating in
the project run ``client'' programs.  Whenever an experiment detects a
candidate burst, the client program makes a connection to the server
machine and sends it an alarm datagram via direct socket
connection. The alarm message contains information about which
experiment observed the burst, along with the time stamp information.
The datagram will be expanded in the future to include information
about the significance and size of the burst.

When the server receives an alarm message from any experiment, it
places the alarm in a queue sorted by UT time, and searches through
all alarm messages in the queue for a coincidence within a given time
window (currently 10 seconds). If there are two or more different
experiments in coincidence, it sends out an alarm.  A test coincidence
server has been set up at the Super-K site in Mozumi, Japan.
Currently, MACRO, Super-K and LVD are connected.  Privacy is
maintained, and security precautions are taken.  Additional servers
can be set up at other sites.

\section*{What do astronomers want?}

What astronomers want from an early supernova alert 
can be summarized by the ``three P's'': ``prompt'', ``pointing''
and ``positive''.  This section describes how SNEWS can address
these ``P's''.

\subsection{``Prompt''}

The alert must be as prompt as possible to catch the early stages of
shock breakout.  All detectors currently in the coincidence
can provide an alert datagram within 30 minutes (worst case, and to be
improved) of the time of the first event in the detector, and in most
cases within only a few minutes. Delays are usually due to buffering at
the detectors.  The coincidence itself and resultant alarm message
take only the time needed for a network connection.  It will be
entirely feasible to have a coincident alarm message produced within
about 15 minutes of the neutrino signal.

\subsection{``Pointing''}

Clearly, the more accurately we can point to a core collapse event
using neutrino information, the more likely it will be that early
light turn on will be observed by astronomers.  Even for the case
when no directional information is available (e.g. for a single
scintillator detector online) it is still useful for astronomers to
know that a gravitational collapse event has occurred. However any
pointing information at all is extremely valuable.  The question of
pointing to the supernova using the neutrino data has been examined
in detail in reference~\cite{Beacom}.

\begin{itemize}
\item \textbf{Asymmetric reactions:} Water Cherenkov detectors can
exploit the neutrino-electron elastic scattering reaction, to point
back to the supernova source.  For a collapse at the center of the
Galaxy, a few hundreds of elastic scattering events are expected in
Super-Kamiokande and tens are expected in SNO.  The recoil electrons follow the
neutrino direction with an opening angle of about 25$^\circ$.  One can
make a rough, optimistic estimate of the pointing resolution from
$\delta \theta~\sim~\frac{25^\circ}{\sqrt{n}}$, where $n$ is the
number of observed events; for 200 elastic scattering events, $\delta
\theta~\sim~2^\circ$.  However, the problem is really that of finding
the center of a peak on top of background, and more realistic
estimates of the resolution yield somewhat worse results.
Reference~\cite{Beacom} estimates a correction factor of 2-3, giving 5
degree pointing for Super-K and 20 degree pointing for SNO.

\item \textbf{Triangulation:} In principle source direction
information can be deduced from the timing of neutrino events at the
different detectors.  Since flight time across the Earth is of order
tens of ms, for successful triangulation the time of the neutrino
pulses at the individual detectors must be tagged to milliseconds or
less.  Since the neutrinos in the pulse are emitted over tens of
seconds, the individual detectors must perform a pulse registration
with limited sampling statistics.  Reference~\cite{Beacom} has studied
the statistical problem in detail, concluding that with the current
generation of detectors (Super-K and SNO), concluding that
triangulation is not promising even in the best case.  There are
additional practical difficulties: a prompt triangulation requires
immediate and complete exchange of event-by-event information, which
is difficult in practice.  However, for a very close supernova, or if
the neutrino pulse comprises unexpectedly sharp features,
triangulation may still be feasible.  Any information is better than
none and the triangulation may at least provide a cross-check of the
elastic scattering pointing.
\end{itemize}

\subsection{``Positive''}

SNEWS must not disseminate any false supernova alarms to the
astronomical community.  One cannot realistically decrease the false
alarm rate to zero, since individual experiments will usually have a
residual rate of false alarms from Poissonian and non-Poissonian
sources; there will then be accidental coincidences between signals
from the individual detectors.  One must weigh the
increased sensitivity from lowering of the alarm thresholds 
(both for the individual experiments and for the coincidence) against
potential waste of resources (and loss of credibility) from issuance
of false alerts.  

\begin{figure}[t] 
\centerline{\epsfig{file=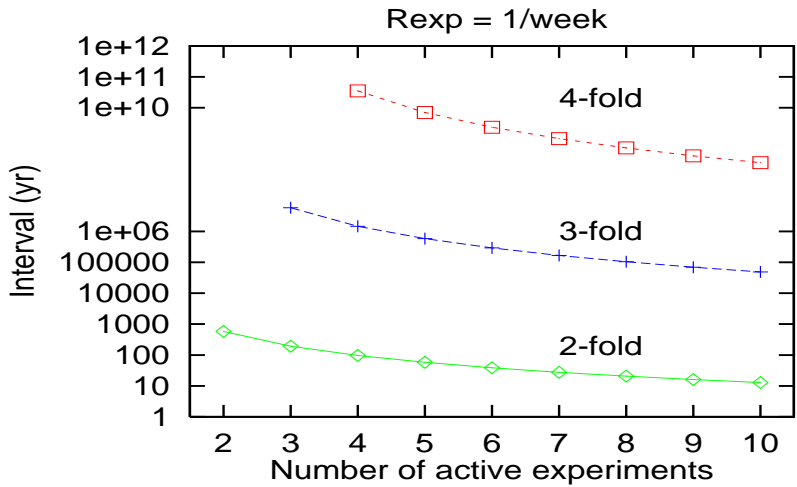,height=2.5in,width=4.5in}}
\vspace{10pt}
\caption{Average time between accidental coincidence 
  alarms between experiments, as a function of number of experiments
  online, for a 10 second coincidence time window, for 2-fold, 3-fold
and 4-fold coincidences.
The assumed individual experiment alarm rate is $R=$1/week.}
  \label{fig:acccoinc1}
\end{figure}

We have chosen the nominal acceptable average false alarm rate to be
one per century.  Assuming equal, constant, uncorrelated
alarm rates for each experiment, and a 10~second coincidence
window, one can calculate the average interval between accidental
alarms for an $n$-fold coincidence of $N$ experiments.
Figure~\ref{fig:acccoinc1} shows the result for an individual
experiment background alarm rate of 1 per week: this rate is
acceptable only if fewer than 4 experiments are online, or if a 3-fold
coincidence if required; otherwise a lower individual experiment rate
is required.  We are also investigating any possibility of
non-Poissonian alarms correlated between experiments.

\section*{The Astronomical Alert}

The astronomical alert from a SNEWS coincidence will be sent out to a
mailing list of interested parties.  In an ideal case, the coincidence
network provides the astronomical community with an event time and an
error box on the sky at which interested observers could point their
instruments.  In a realistic case, the error box, which is dependent
on the location of the supernova and experiments which are online, may
be very large.  However, members of the mailing list with wide-angle
viewing capability (satellites, small telescopes and amateurs) should
be able to pinpoint an optical event quickly.

\section*{An Additional Role for SNEWS?}

So far, SNEWS has been intended to provide an early warning for
astronomers.  However, it has another potential role.  If a core
collapse supernova happens in the Galaxy, it will be an unprecedented
opportunity for science, and all possible data -- neutrinos,
electromagnetic, gravitational waves, perhaps other kinds --
would be extremely valuable.  But many detectors which are capable of
providing useful information are not necessarily capable of triggering
themselves on a supernova burst and may not be continuously archiving
information.  They may be noisy and/or may not know what kind of signal to
look for from a supernova.  Some examples of detectors in this
category would be: some of the long string detectors
(ANTARES, Baikal), gravitational wave detectors (if not all data is
archived), and surface neutrino-sensitive detectors with a high rate of cosmic
ray background.  The SNEWS neutrino coincidence will be a high
confidence indication that a supernova has occurred.  Noisy SN
detectors could therefore arrange to use the SNEWS coincidence as an
input -- they could set up a buffering system to record data (for
hours or days, depending on resources available) that would routinely
be overwritten, but which could be saved to permanent storage in the
case of a SNEWS coincidence.  This approach would greatly enrich the
world's supernova data sample.

\section*{Current Status and Future}

Currently, a test coincidence server is running at the Super-K site in
Mozumi.  Three experiments are online (Super-K, MACRO and LVD),
sending alarm datagrams in test mode.  SNO and AMANDA are expected to
join within about 6 months.  There is no automated alert to
astronomers yet; we expect the automated alert to be activated after a
test period.

\section*{Acknowledgments}

The author wishes to thank all the members of the SNEWS inter-experiment
working group (in particular Alec Habig) and John Beacom.


\begin{references}
\bibitem{Burrows}Burrows A. {\it et al.},  {\it Phys. Rev.} {\bf
D45}, 3362 (1992).
\bibitem{snews} The SNEWS web page, \texttt{http:/hep.bu.edu/$\sim$snnet/}.
\bibitem{Beacom} Beacom J. F. and P. Vogel, {\it Phys. Rev.} {\bf D60} 033007 (1999).  astro-ph/9811350
\bibitem{HST} J. Bahcall (P.I.), "Observing the next nearby supernova", HST proposal 8404.
\end{references}
\end{document}